\documentclass[fleqn,10pt,twocolumn, final]{SPIEmain}
\usepackage[utf8]{inputenc}
\usepackage[T1]{fontenc}
\usepackage{graphicx}
\usepackage{siunitx}
\usepackage{float}
\usepackage{authblk}
\usepackage{lettrine}
\usepackage{soul}
\usepackage{pdfpages}
\usepackage{orcidlink}
\usepackage{anyfontsize}
\usepackage{siunitx}
\DeclareSIUnit{\riu}{RIU}
\soulregister\cite7
\soulregister\ref7 

\begin{document}

\title{Thermo-optic dynamics of effective epsilon-near-zero media}

\author[1,5,$\dagger$,*]{Jiaye Wu\orcidlink{0000-0002-0650-1274}}
\author[2,$\dagger$]{Xuanyi Liu\orcidlink{0000-0001-7542-3882}}
\author[3]{Marco Clementi\orcidlink{0000-0003-4034-4337}}
\author[2]{Shuang Qiu\orcidlink{0009-0004-7332-3238}}
\author[2]{Limin Lin\orcidlink{0000-0002-8275-5727}}
\author[2,4,*]{Zhang-Kai Zhou\orcidlink{0000-0002-4341-3097}}
\author[1,*]{Camille-Sophie Br\`{e}s\orcidlink{0000-0003-2804-1675}}

\affil[1]{  \'{E}cole Polytechnique F\'{e}d\'{e}rale de Lausanne (EPFL), Photonic Systems Laboratory (PHOSL), STI-IEM, Station 11, Lausanne CH-1015, Switzerland.}
\affil[2]{ State Key Laboratory of Optoelectronic Materials and Technologies, School of Physics, Sun Yat-sen University, Guangzhou 510275, China.}
\affil[3]{ Dipartimento di Fisica “A. Volta", Università di Pavia, Via A. Bassi 6, 27100 Pavia, Italy.}
\affil[4]{ Quantum Science Center of Guangdong-Hong Kong-Macao Greater Bay Area, Shenzhen, China.}
\affil[5]{ City University of Hong Kong (Dongguan), Dongguan 523808, China.}
\affil[$\dagger$]{ These authors contribute equally to this work.}
\affil[*]{ Corresponding authors. Email: jiaye.wu@epfl.ch; jiaye.wu@cityu-dg.edu.cn; zhouzhk@mail.sysu.edu.cn; camille.bres@epfl.ch}

\begin{abstract}
\textbf{Abstract}

Epsilon-near-zero (ENZ) photonic media exhibit extreme optical dispersion that enables unconventional light-matter interactions and enhanced optical nonlinearities.
Recent studies suggested that thermo-optic effects, traditionally regarded as slow and secondary, can be strongly modified under the ENZ condition.
Here we establish thermo-optic reconfiguration of effective media as a unified physical framework to describe both static and transient thermo-optic phenomena in ENZ systems.
Using an effective medium operating in the visible spectral range, we experimentally demonstrate that temperature variation, whether under thermal equilibrium or transient excitation, reconfigures the constitutive parameters defining the ENZ condition, giving rise to pronounced linear and nonlinear optical responses.
At thermal equilibrium, this reconfiguration manifests itself as static ENZ wavelength shift with an unprecedentedly large thermal-spectral modulation rate and an effective thermo-optic coefficient on the order of $10^{-1} $\si{\per\kelvin}.
Under ultrafast excitation, we observe a picosecond-scale thermo-optic nonlinear response induced by transient heating.
This response can be consistently interpreted as a time-dependent reconfiguration of the effective ENZ medium, corresponding to a transient evolution of its optical parameters.
By reframing thermo-optic effects as a process of static and dynamic reconfiguration of effective media, this work provides a unified perspective that bridges thermo-optic physics, effective-medium theory, and time-varying photonics.

\end{abstract}

\twocolumn
\captionsetup[figure]{labelfont={bf},name={Fig.},labelsep=none}
\maketitle

 \fontsize{10pt}{10pt}\selectfont

\section{Introduction}

\noindent The intriguing field of near-zero-index (NZI) photonic media, and in particular epsilon-near-zero (ENZ) photonics, has attracted extensive scientific and engineering interest owing to the unconventional light-matter interactions enabled by vanishing refractive index or permittivity \cite{Niu2018,Reshef2019,Kinsey2019,Wu2021a,Xie2025}.
In such media, ordinary physical principles give rise to counterintuitive phenomena, including phase tunneling \cite{Silveirinha2006}, efficient harmonic and terahertz generation \cite{Capretti20151,Luk2015,Yang2019,Tian2021,Tirole2024,Jia2021,Minerbi2022}, frequency translation via time-varying or intracavity ENZ interfaces \cite{Khurgin2020,Zhou2020,Zhang2020,Wu2022lpr,Wu2025}, as well as pronounced field enhancement and nonlinear optical responses \cite{Campione2013,Alam2016,Alam2018,Deng2020,Khurgin2020a,Wu2024}.
These properties have positioned ENZ platforms as a fertile ground for exploring extreme optical dispersion, nonlinearities, and reconfigurable photonic functionalities.

Thermo-optic effects, although ubiquitous in optical materials, have traditionally been regarded as slow perturbations and are therefore often neglected in the context of ultrafast or high-frequency photonics \cite{Boyd2020}.
Recent studies have shown, however, that this conventional perception does not necessarily hold in ENZ media.
Owing to the singular dispersion near the ENZ condition, both linear and nonlinear thermo-optic responses can be substantially enhanced, while the characteristic temporal thresholds associated with thermal processes can be pushed into the sub-picosecond regime \cite{Wu2024}.
Moreover, many optical-frequency ENZ materials are compatible with complementary metal-oxide-semiconductor (CMOS) technology \cite{Wang2024}, rendering thermo-optic effects particularly relevant for on-chip ENZ photonics, where electro-optic integration and packaging inevitably introduce heat generation.

To date, thermo-optic phenomena in nanophotonic systems have been extensively investigated across a wide range of material platforms, including plasmonic and CMOS-compatible materials such as indium tin oxide (ITO), primarily through theoretical analyses and platform-specific demonstrations \cite{Khurgin2015,Khurgin2020a,Iadanza2020,Clementi2021,Sarkar2023}.
However, the explicit role of thermo-optic effects under the ENZ condition, where the vanishing permittivity fundamentally reshapes optical dispersion and light-matter interaction, has only recently begun to be experimentally explored \cite{Wu2024}.
As a result, the temporal and spectral characteristics  of thermo-optic responses in ENZ media remain largely uncharted.

\begin{figure*}[!ht]
	\footnotesize
	\includegraphics[width=1\linewidth]{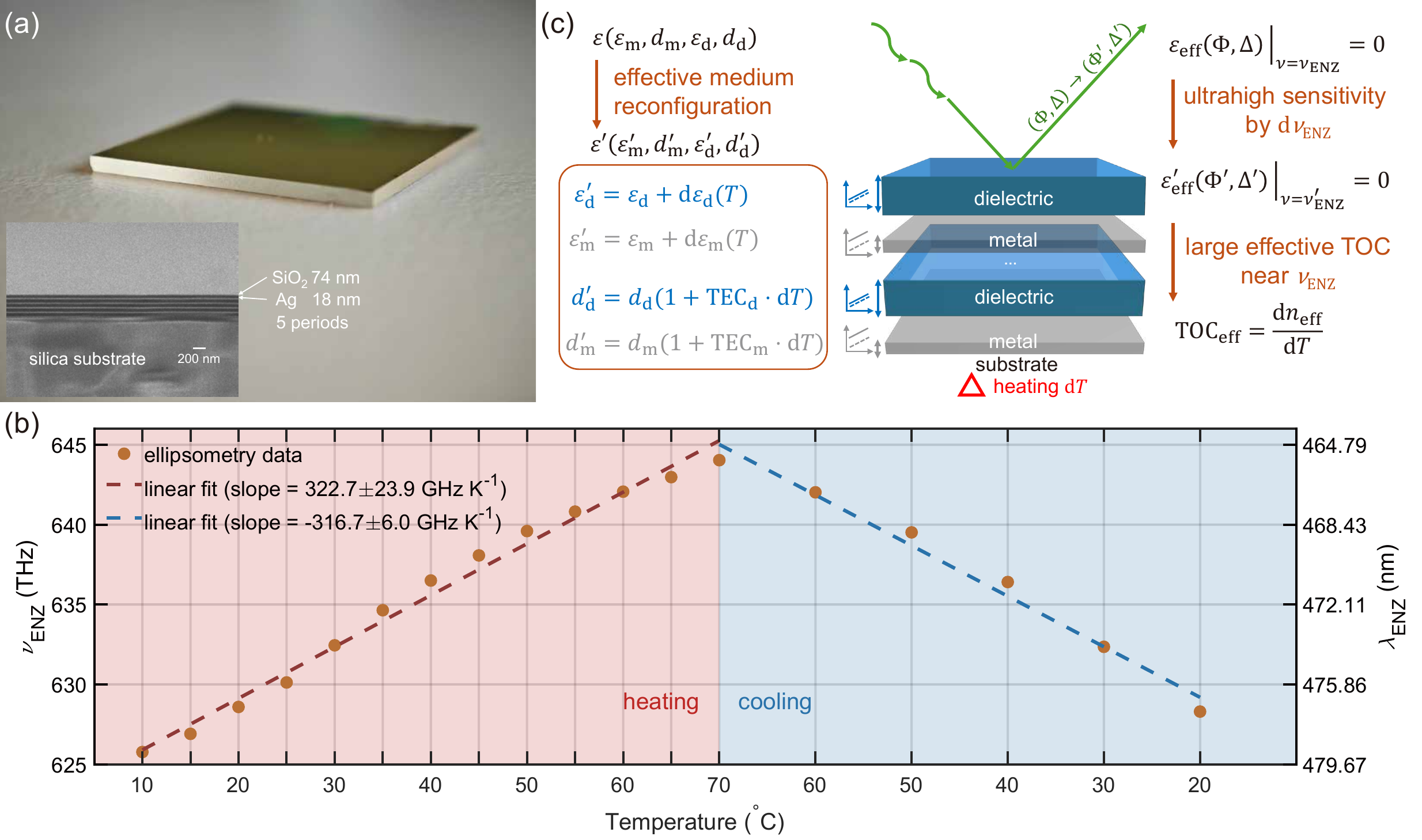} 
	\caption{{. High sensitivity temperature detection with ENZ frequency.} 
    (a) Photograph and scanning electron microscope (SEM) image of the multilayered effective ENZ medium.
    The original SEM image with details is shown in Supplementary Figure 2. 
    The quartz glass substrate of the sample has a dimension of $20\times20\times1.1$ \si{\milli\metre}.
    (b) Measured variation of the ENZ frequency with temperature.
    (c) Schematic diagram of the effective-medium reconfiguration and the mechanism for ENZ-induced large effective TOC.
    Under the thermo-optic excitation (equilibrium or transient), the configuration of effective medium $\varepsilon(\varepsilon_{\rm m},d_{\rm m},\varepsilon_{\rm d},d_{\rm d})$ becomes $\varepsilon'(\varepsilon'_{\rm m},d'_{\rm m},\varepsilon'_{\rm d},d'_{\rm d})$ due to different extent of material-specific thermo-optic and thermal expansion coefficient changes.
    }
	\label{f1}
\end{figure*}

In an effective-medium ENZ platform, the optical response is jointly determined by the permittivities and geometrical parameters of its constituent materials.
When the system is subjected to heating, either under thermal equilibrium or during transient excitation, the permittivities of the dielectric and metallic components evolve through the thermo-optic effect, while their thicknesses simultaneously change through thermal expansion, generally with different magnitudes and signs.
Consequently, temperature variation does not merely perturb a single material parameter but instead reconfigures the combination of parameters that define the effective medium itself, giving rise to a new ENZ state.
We refer to this process as thermo-optic reconfiguration of the effective medium.
At thermal equilibrium, it manifests as a static reconfiguration characterized by a recoverable shift of the ENZ condition; under ultrafast excitation, it naturally extends to a time-varying reconfiguration, in which the effective ENZ medium evolves dynamically in time.

In this work, we experimentally demonstrate both the static and dynamic thermo-optic reconfiguration of a CMOS-compatible effective ENZ medium operating in the visible spectral range.
By employing a metal-dielectric multilayer architecture, we first show a static thermo-optic ENZ reconfiguration at thermal equilibrium, exhibiting a spectral modulation trend distinct from previously reported ENZ platforms and a markedly enhanced thermal-spectral response.
Owing to the oscillatory dispersion profile inherent to the effective ENZ medium, we further observe a giant enhancement of the effective thermo-optic coefficient, reaching values on the order of $10^{-1} $\si{\per\kelvin}.
Finally, using an ultrafast pump-probe experiment, we observe in real-time a picosecond-scale thermo-optic nonlinear response induced by transient heating.
Supported by effective-medium modeling and numerical simulations, this fast response can be consistently interpreted as a dynamic, time-dependent reconfiguration of the effective ENZ medium, corresponding to a transient evolution of its constitutive parameters.

The concept of thermo-optic reconfiguration establishes a direct link between effective-medium physics and functional device behavior.
The pronounced sensitivity of the ENZ condition to temperature-induced reconfiguration enables static temperature sensing with a sensitivity comparable to state-of-the-art organic materials, while surpassing all previously reported CMOS-compatible inorganic photonic platforms, to our knowledge.
More importantly, the picosecond-scale thermo-optic response arising from dynamic reconfiguration allows the detection of transient heat sources that are fundamentally inaccessible to conventional thermal sensors.
Such capability is particularly relevant for large-scale-integrated (LSI) photonic circuits, where ultrafast optical pulse leakage or localized overheating in defective or stressed components can occur on timescales far shorter than those addressed by existing thermal monitoring technologies, offering a powerful tool for thermal risk management in advanced LSI photonic systems.

\section{Static thermo-optic reconfiguration of the effective ENZ medium}

\noindent \noindent The zero-crossing frequency, $\nu_{\rm ENZ}$, denotes the transition of the ENZ medium between dielectric-like and metal-like behaviors.
It usually implies the emergence and existence of ENZ-induced optical effects within its proximity, while the ENZ frequency itself is less focused on.
When designing an effective ENZ device using the effective medium theorem (EMT) \cite{gaylordZeroreflectivityHighSpatialfrequency1986}, the spectral location of $\nu_{\rm ENZ}$ (or equivalently, $\lambda_{\rm ENZ}$) is the design goal.

To achieve ENZ effects in the visible range, we designed and fabricated a 5-period nanophotonic thin film device with each bilayer pair consisting of 74-nm amorphous silica (SiO$_2$) and 18-nm silver (Ag).
A photograph and a scanning electron microscope (SEM) image of the device are shown in Fig.~\ref{f1}(a) with details provided in Supplementary Note 2.
The layer thicknesses are chosen so that important wavelengths in the green, such as \SI{516.67}{\nano\metre} (third harmonic of \SI{1550}{\nano\metre}) and \SI{532}{\nano\metre} (second harmonic of \SI{1064}{\nano\metre}), are covered by the ENZ enhancement (for example, see the nonlinearity enhancement region of Figs.~\ref{f3}(b--c)) after taking into account a possible 5\% fabrication error.
The permittivity of the device is described by the effective medium theorem (EMT) and modeled by the Maxwell-Garnett equations \cite{Niu2018,Wu2021a}:

\begin{equation}
    \varepsilon_{\parallel} = \frac{\varepsilon_{\rm d}d_{\rm d} + \varepsilon_{\rm m}d_{\rm m}}{d_{\rm d} + d_{\rm m}}, \varepsilon_{\perp} = \frac{\varepsilon_{\rm d}\varepsilon_{\rm m}(d_{\rm d} + d_{\rm m})}{\varepsilon_{\rm d}d_{\rm m} + \varepsilon_{\rm m}d_{\rm d}},
    \label{e1}
\end{equation}

\noindent where the subscripts $\parallel$ and $\perp$ are with respect to the normal direction of the structure, and $\rm d, m$ denote dielectric and metal. With proper combinations of the dispersion and thicknesses of the two materials, $\varepsilon_{\parallel} = 0$ can be achieved at certain wavelengths, and in our case, the theoretical $\lambda_{\rm ENZ} = 483.8$ \si{\nano\metre} (see Supplementary Note 1).

In the effective medium, the anisotropic permittivities exist in the form of tensors, and are experienced by light as an effective response.
A custom-built ellipsometry-compatible sample stage with precise temperature feedback control allows the measurement of $\nu_{\rm ENZ}$ sensitivity and the observation of the thermo-optic effects under an ellipsometer without an external pump source (see Supplementary Note 4).
Here, the effective permittivity of the metamaterial as a whole is determined by a pseudo transform (see Supplementary Note 1) based purely on the ellipsometry data $(\Phi, \Delta)$:

\begin{equation}
    \varepsilon_{\rm eff} = \sin^2 \theta_{\rm i} \left[ 1+ \tan^2 \theta_{\rm i} \left( \frac{1 - \rho}{1 + \rho}\right)^2 \right],\quad \rho = \tan(\Phi)e^{i\Delta}.
    \label{e2}
\end{equation}

\begin{figure*}[th]
	\centering
	\includegraphics[width=1\linewidth]{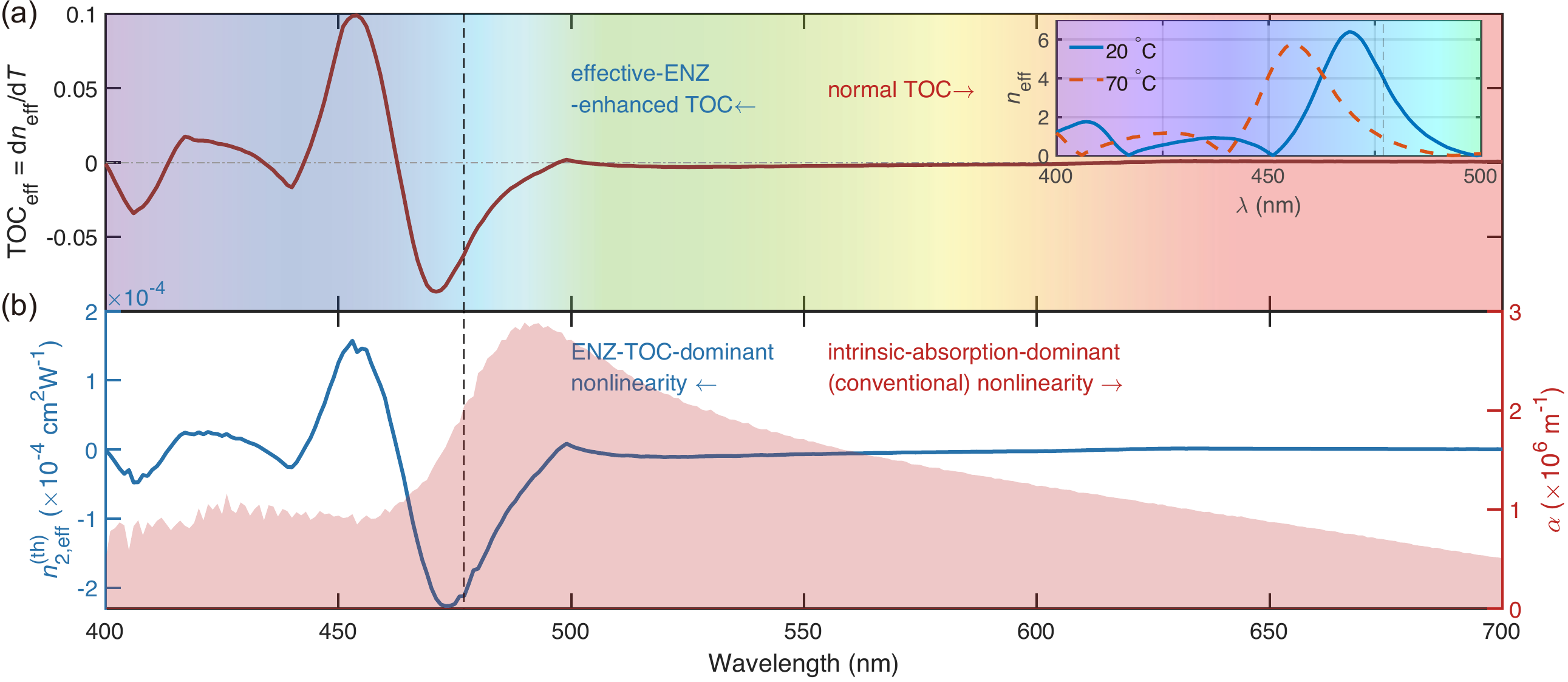} 
	\caption{{. Enhancement of thermo-optic effects under ENZ conditions.} 
    (a) Experimental thermo-optic coefficient (${\rm TOC_{eff}}$) calculated from ellipsometric curves at different temperatures.
    The inset shows a comparison of the $n_{\rm eff}$ curves calculated at low and high temperature, highlighting the origin of the unprecedentedly high ${\rm TOC_{eff}}$.
    (b) The evolution of effective thermo-optic nonlinearity ($n_{\rm 2,eff}^{\rm(th)}$) and loss ($\alpha$) from ENZ-enhanced region to intrinsic-absorption-dominant region at longer wavelengths.
    The dashed vertical line indicates $\lambda_{\rm ENZ}$ at \SI{20}{\celsius}.
    The long-wavelength side of the spectra marks the thermo-optic behavior of the effective medium far away from ENZ enhancement, where ${\rm TOC_{eff}}$ and $n_{\rm 2,eff}^{\rm(th)}$ drops to the level of conventional optical materials.
    }
	\label{f2}
\end{figure*}

\noindent where $\Phi$ is the ratio between the reflected amplitude of $s$- and $p$-polarizations, and $\Delta$ is their phase difference.
The benefit of this method is that it is ellipsometry-model-free, and it greatly reduces the potential errors induced by inappropriate model selection/fitting settings and local (false) minimum mean square error (MSE).
The obtained effective $\lambda_{\rm ENZ}\sim$ \SI{478}{\nano\metre} at \SI{65}{\degree} incidence is quite close to the theoretical design (assuming normal incidence), also indicating a good angular uniformity.

In this work, we measure the temperature dependence of $\nu_{\rm ENZ}$ in our metamaterial from \SI{10}{\celsius} to \SI{70}{\celsius} with \SI{5}{\celsius} interval at thermal equilibrium states (see Supplementary Note 4), and then gradually cool the sample down to \SI{20}{\celsius}.
The maximum \SI{70}{\celsius} is limited by the temperature controller used.
The extracted $T$-$\nu_{\rm ENZ}$ curve is depicted in Fig.~\ref{f1}(b). 

In the previous study of the linear thermo-optic ENZ effect (temperature-induced $\nu_{\rm ENZ}$ shift) \cite{Wu2024}, $\nu_{\rm ENZ}$ of ENZ ITO was found to decrease with rising temperature under the reversible, sub-annealing temperatures ($\le$~\SI{100}{\celsius}). 
An adapted grain-boundary barrier model \cite{Morris1980} and a vacuum treatment experiment were used to explain this effect, demonstrating that oxygen diffusion into the lattice and the occupation of trap states during heat-up were the main causes of this increase. 
This also suggests that this effect is material-specific and may not be applicable to other ENZ realization platforms.
We can see that the variation trend in Fig.~\ref{f1}(b) is opposite to that previously reported in ITO: here $\nu_{\rm ENZ}$ \textit{increases} rapidly with temperature.
Although $\nu_{\rm ENZ}$ can also increase with temperature in ENZ transparent conducting oxides like ITO under annealing processes, in these cases the changes are permanent and irreversible \cite{Huang2023, Xian2019, Wang2020,Wu2021b}, and they cannot be regarded as thermo-optic effects.

The thermo-optic behavior of our effective-medium ENZ metamaterial can be explained by considering the thermal reconfiguration of $(\varepsilon, d)$ parameter pairs in Eq.~\eqref{e1}.
In conventional homogeneous optical materials, the thermo-optic effect is induced by a geometric effect causing the change of lattice constant upon heating and a thermo-refractive effect involving the thermal excitation of the carriers, bandgap shrinking, diffusive currents, \textit{etc} \cite{Carmon2004, Iadanza2020, Clementi2021}.
However, in the effective ENZ medium, this is a cascaded process.

The metal and dielectric layers have contrasting thermal expansion coefficients \cite{Bachmann1988,Bogatyrenko2021} and temperature-dependencies of permittivity \cite{Ferrera2019} $\varepsilon(\lambda, T)$.
These two factors contribute to the system both locally, within the layers and globally by reconfiguring the effective-medium superlattice.
In our case, silver is known to have a much higher thermal expansion coefficient \cite{Bogatyrenko2021} than silica \cite{Bachmann1988}, which indicates that in Eq.~\eqref{e1} the contribution of the negative permittivity \cite{Babar2015} now becomes larger.
Considering the thermal expansion effect alone, the 18-nm Ag and 74-nm SiO$_2$ will become \SI{18.0243}{\nano\metre} and \SI{74.0018}{\nano\metre}, respectively, under $\Delta T = 50$ \si{\kelvin}.
While these changes might seem trivial, but when considered in the EMT equation Eq.~\eqref{e1}, the original $\nu_{\rm ENZ(\parallel,theoretical)} = 483.8$ \si{\nano\metre} will be blueshifted to \SI{483.6}{\nano\metre}.
The 0.2-nm difference translates to \SI{256.38}{\giga\hertz}, which already yields a thermal-spectral modulation rate of \SI{5.38}{\giga\hertz\per\kelvin}, which is already on par with that in ENZ ITOs \cite{Wu2024}.

At the nanoscale, the absolute value of $\varepsilon_{\rm Ag}(\lambda, T)$ also becomes larger with rising temperature \cite{Ferrera2019}, which further increases the contribution of the negative permittivity in Eq.~\eqref{e1}.
Therefore, the ENZ condition can only be fulfilled at a shorter wavelength (blueshift), where the positive permittivity \cite{Franta2016} of silica is higher.
With $\Delta T = 50$ \si{\kelvin}, silver has a ${\rm d}\varepsilon/{\rm d}T$ value \cite{Ferrera2019} at $-10^{-3}$ \si{\per\kelvin}, which is 2 orders of magnitude larger than that of silica \cite{Rego2023}, elevating the thermal-spectral modulation rate to the level of $\sim190$ \si{\giga\hertz\per\kelvin} by linear estimation.
These mechanisms are illustrated in Fig.~\ref{f1}(c).

As seen from the ellipsometery results, these combined effects enhance the thermal-spectral modulation rate (temperature sensitivity) at $\nu_{\rm ENZ}$ to an unprecedented $322.7\pm23.9$ \si{\giga\hertz\per\kelvin}, which is 2 orders of magnitude higher than the records in ENZ ITOs \cite{Wu2024} at \num{-8.45} to \SI{-6.45}{\giga\hertz\per\kelvin}.
From the fit uncertainty, assessed at 95\% confidence, we infer a resolution of the metamaterial as a temperature sensor as low as \SI{7.41e-2}{\kelvin}.

It is important to emphasize that the physical mechanisms governing the thermo-optic dynamics in the effective ENZ medium differ fundamentally from those in homogeneous ENZ materials, such as degenerate semiconductors like the ENZ ITO. 
In homogeneous ENZ films, the relatively lower electron density restricts the energy transfer to the lattice, typically resulting in a minimal lattice temperature rise. 
In stark contrast, the high electron density of the constituent metal (Ag) in our stratified platform ensures a massive energy transfer during electron-lattice relaxation, driving a substantial macroscopic temperature rise. 
Furthermore, our multilayer architecture inherently introduces strong thermal anisotropy, featuring high in-plane thermal conductivity alongside restricted out-of-plane heat dissipation. 
This unique spatial thermal confinement is a distinct advantage over isotropic homogeneous films in applications such as thermo-optic temperature sensing.

\section{Large enhancement of effective TOC at the $10^{-1}$ K$^{-1}$-level and the resulting thermo-optic nonlinearity} 

\noindent A direct consequence of the ultra-large thermal-spectral modulation rate is the existence of a record-breaking ${\rm TOC_{eff}}$, which holds significant implications for the thermo-optic-nonlinearity-related analysis in this work. 
The linear TOC is defined by refractive index change with temperature: ${\rm TOC_{eff}} = {\rm d}n/{\rm d}T$.
Within the ENZ region, a near-zero $\varepsilon$ yields an enhanced TOC -- and in the anisotropic effective ENZ medium's case, an enhanced ${\rm TOC_{eff}}$, which can be approximated as \cite{Wu2024}:

\begin{equation}
    \Delta n_{\rm eff} \overset{\Delta\varepsilon_{\rm i} \to 0}{\approx} \frac{{\Delta {\varepsilon_{\rm r}}}}{{\sqrt {{\varepsilon_{\rm r}} + \Delta {\varepsilon_{\rm r}}}  + \sqrt {{\varepsilon_{\rm r}}} }} \approx \frac{\Delta\varepsilon_{\rm r}}{2\sqrt {\varepsilon_{\rm r}}},
\end{equation}

\noindent where $\varepsilon_{\rm r} = \Re(\varepsilon_{\rm eff})$ and $\varepsilon_{\rm i} = \Im(\varepsilon_{\rm eff})$ are the real and imaginary parts of the effective permittivity.
By comparing the ellipsometry data of the sample at high and low temperatures, the spectrum of ${\rm TOC_{eff}}$ is illustrated in Fig.~\ref{f2}(a).
Note that ${\rm TOC_{eff}}$ is an effective value ``seen'' by light (see Supplementary Note 1), and it is not an intrinsic property of the bulk or any of its components.

The large thermal-spectral modulation rate gives rise to a great spectral shift of $n_{\rm eff}$ (Eq.~\eqref{e2}) through $n_{\rm eff} \propto \sqrt{\varepsilon_{\rm eff}}$ as shown in the inset of Fig.~\ref{f2}(a), where a spectral offset between $n_{\rm eff}$ peaks and valleys emerge.
Therefore, for the first time in ENZ media, we record the coexistence of positive ($n$ increases with rising $T$) and negative ($n$ decreases with rising $T$) ${\rm TOC_{eff}}$ in the enhancement region.
Due to the $n_{\rm eff}$ peak-valley offset, ${\rm TOC_{eff}}$ exhibits an unprecedented \textit{effective} value as high as $\sim \pm 10^{-1}$ \si{\per\kelvin}, significantly higher than any conventional organic and inorganic bulk materials reported (see Supplementary Tables 2 and 3).

\begin{figure*}[!h]
	\centering
	\includegraphics[width=0.95\linewidth]{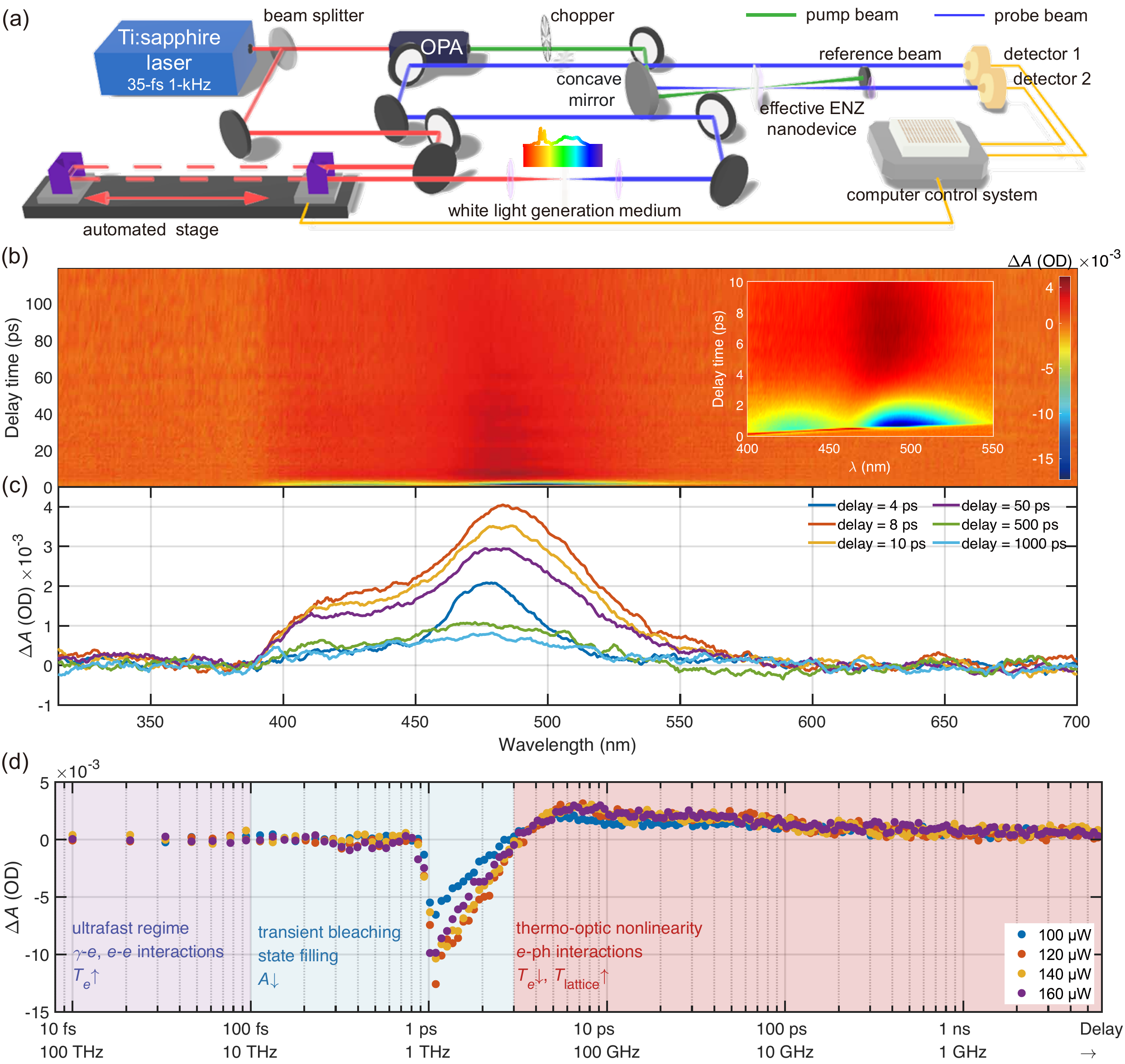} 
	\caption{{. Ultrafast pump-probe experiments at the visible range.} 
    (a) Experimental scheme of the setup.
    (b) Spectral-temporal diagram of the absorption variation, clearly showing the main peak (enhancement range) of the thermo-optic effects. 
    The inset shows the real-time ultrafast dynamics in the first \SI{10}{\pico\second}.
    (c) Pump-probe spectra at selected delays. 
    (d) Ultrafast dynamics of the whole processes in the pump-probe light-matter interactions at different pump powers with $\lambda_{\rm pump} = 475$ \si{\nano\metre} near $\lambda_{\rm ENZ}$.
    }
	\label{f3}
\end{figure*}

\begin{figure*}[!h]
	\centering
	\includegraphics[width=1\linewidth]{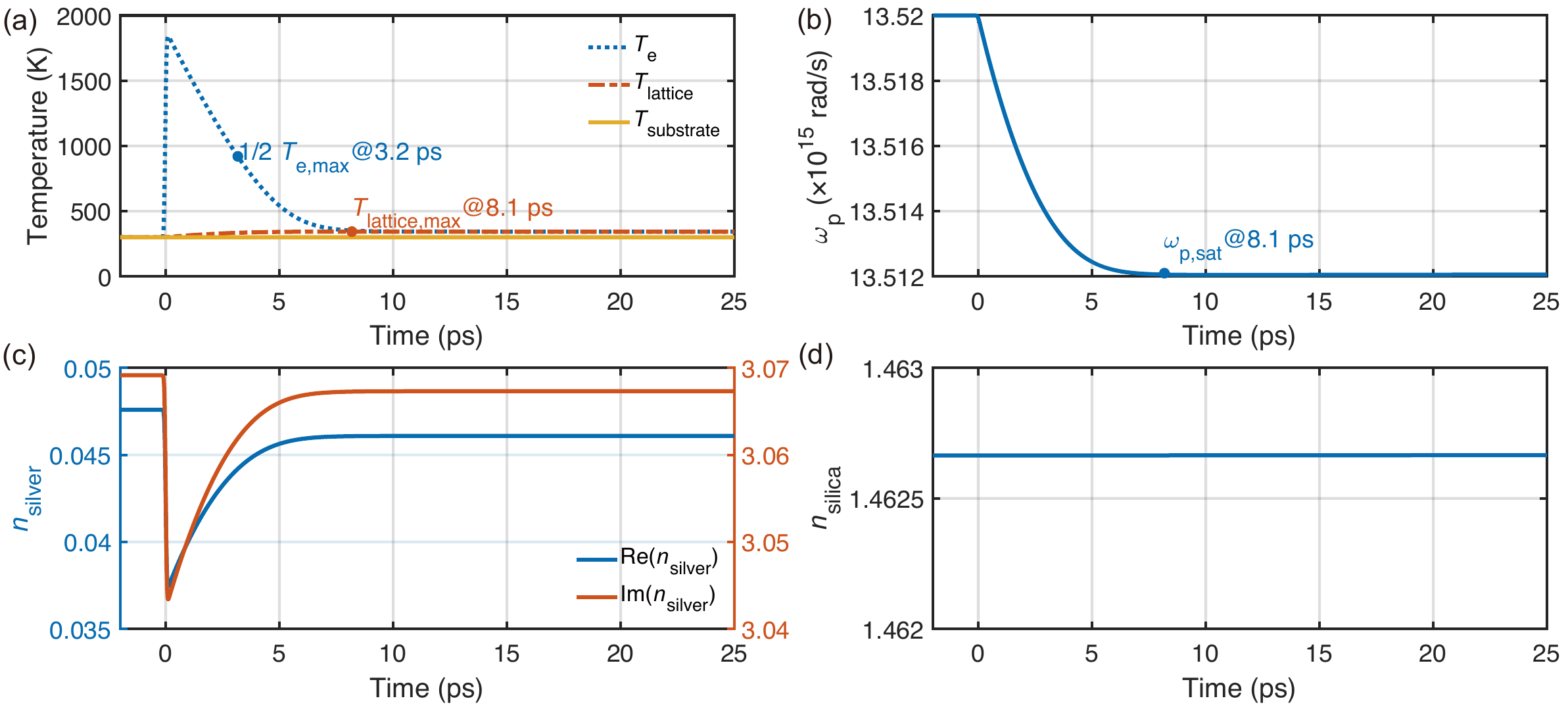} 
	\caption{{. TTM simulation of picosecond-grade thermo-optic nonlinearity response.} 
    (a) Electron, lattice, and substrate temperature evolution with time near $\lambda_{\rm ENZ}$.
    $T_{\rm e}$ drops to half of its maximum intensity at \SI{3.2}{\pico\second} where the thermo-optic nonlinearity becomes visible in Figs.~\ref{f3}(c) and (d).
    $T_{\rm lattice}$ reaches its maximum at \SI{8.1}{\pico\second} which agrees well with experiments.
    (b) The evolution of thermo-expansion-corrected plasma frequency with time near $\lambda_{\rm ENZ}$.
    $\omega_{\rm p}$ also saturates near \SI{8.1}{\pico\second}.
    The evolution of refractive indices of the component materials (c) silver and (d) silica near $\lambda_{\rm ENZ}$ with time, showing a time-varying dynamic reconfiguration of the effective ENZ medium.
    }
	\label{f4}
\end{figure*}

At wavelengths farther away from $\lambda_{\rm ENZ}$, the influence of ENZ dispersion becomes much weaker.
Near $\lambda \approx 550$ \si{\nano\metre}, $\rm{TOC_{eff}}$ drops to an ordinary value as compared to silver nanoparticles \cite{Karimzadeh2010} in the range of $10^{-4}$ \si{\per\kelvin} to $10^{-6}$ \si{\per\kelvin}.

Like the ENZ-enhanced, free-carrier-generation-induced effective Kerr effect \cite{Alam2016} where absorptive loss is the physical origin of large nonlinearity \cite{Secondo2020}, the thermo-optic nonlinearity is driven by both the TOC and the absorption \cite{Boyd2020}:

\begin{equation}
    n_{\rm 2,eff}^{\rm (th)} = \frac{{\rm d}n_{\rm eff}}{{\rm d}T} \cdot \frac{\alpha r^2}{\kappa_{\rm eff}} = {\rm TOC_{eff}} \cdot \frac{\alpha r^2}{\kappa_{\rm eff}},
    \label{e4}
\end{equation}

\noindent where $n_{\rm 2,eff}^{\rm (th)}$ is the effective thermo-optic equivalent of the Kerr nonlinear-index coefficient $n_2^{\rm (el)}$, which modifies $n$ by $n = n_0 + n_{\rm 2,eff}^{\rm (th)} I$, with $I$ being the light intensity and $n_0$ the linear refractive index.
$\alpha$ is the propagation loss (\si{\per\metre}) calculated from the measured transmission and reflection spectra (see Supplementary Note 3) and presented in Fig.~\ref{f2}(b) in red.
$r$ is the laser spot radius, and here we take a typical value \cite{Wu2023} for a transverse mode size in integrated photonics $r = 0.5$ \si{\micro\metre}.
$\kappa_{\rm eff}$ is the weighted heat conductivity, calculated \cite{Zhu2018,Chen2021} to be \SI{1.49}{\watt\per\metre\per\kelvin} (see Supplementary Note 4).
The resulting calculated spectrum of $n_{\rm 2,eff}^{\rm (th)}$ is shown in Fig.~\ref{f2}(b) (blue solid line).
Note that in textbook\cite{Boyd2020} Eq.~\ref{e4} is deduced for continuous wave (CW) only, it holds true for pulses if the subwavelength optical structure is also ``sub-temporal-width'', meaning that the time needed for light transmission is shorter than the temporal width of the pulse.

From comparison with Fig. \ref{f2}(a), it is evident that the envelope of the $n_{\rm 2,eff}^{\rm (th)}$ is largely influenced by the shape of the ${\rm TOC_{eff}}$ curve, while both the positive and negative peaks reside within the ENZ-induced absorption enhancement region.
The existence of ENZ-enhanced $n_{\rm 2,eff}^{\rm (th)}$ peaks is further verified by the pump-probe spectra shown in Fig.~\ref{f3}(c) in the form of a positive change in absorption ($\Delta A$) (see next section).
The peak values of $n_{\rm 2,eff}^{\rm (th)}$ are also 3 orders of magnitude higher than the ENZ-enhanced $n_{\rm 2}^{\rm (th)}$ in ENZ ITOs \cite{Wu2024}.
Away from the ENZ region, the thermo-optic nonlinearity is dominated by intrinsic absorption, which becomes even and uniform.
It is worth emphasizing that $n_{\rm 2,eff}^{\rm (th)}$ has its limitations \cite{Boyd2020}, especially in the ultrafast domain. The use of it here is to show the quantitative magnitude of ENZ-induced enhancement and the fact that its enhancement spectral range aligns well with the independent pump-probe experiment demonstrated in the next section.

\section{Dynamic reconfiguration of effective medium and picosecond-scale thermo-optic nonlinear response} 

Various state-of-the-art theoretical studies predicted that the thermo-optic response (onset/activation) can be accelerated by down-scaling the thermal volume \cite{Khurgin2015, Iadanza2020, Clementi2021}, especially in plasmonic ENZ materials like ITO \cite{Khurgin2020a, Sarkar2023}, aiming to reach the ultrafast, attosecond to picosecond, timescales \cite{Khurgin2020a, Wu2024}. 
The thermo-optic dynamics of a single nano-object has been studied \cite{Rouxel2020UltrafastThermooptical, Rouxel2021ElectronLattice}, and phenomena such as thermally-induced transmission variations \cite{Larciprete2002ThermallyInduced} and optical switching \cite{Larciprete2005OpticalSwitching} have been considered and investigated.
For non-plasmonic effective ENZ metamaterials, such as those considered in this work, the thermo-optic response could additionally be further enhanced.

To observe and verify the real-time dynamics of the fast thermo-optic nonlinearity, we carried out an ultrafast pump-probe experiment at the visible range, with a schematic diagram of the setup shown in Fig.~\ref{f3}(a) (also see Supplementary Note 6). 

Figure~\ref{f3}(b) illustrates the spectral-temporal diagram of the ultrafast pulse excitation by displaying the change in absorption ($\Delta A$) with respect to the background signal before the pump pulse reaches the sample. 
A darker red shade covers the range from \SIrange{400}{550}{\nano\metre}, which is exactly the spectral range where an ENZ-enhanced $n_{\rm 2,eff}^{\rm (th)}$ is found in Figs.~\ref{f2}(b) and c.
Additional pump wavelength detuning experiments also confirm the local enhancement of the thermo-optic nonlinearity (see Supplementary Note 6).

More details on the temporal dynamics can be observed in Fig.~\ref{f3}(d).
At a temporal scale comparable to the pulse width $\sim$~\SI{100}{\femto\second}, a small increase of $\Delta A$ is observed.
The Kerr nonlinearity takes place and the dominant interactions are photon-electron ($\gamma$-e) and electron-electron (e-e), where the effective electron temperature increases.
At this stage, the lattice is still ``cold''.
The actual pulse duration, after being inevitably broadened by optics on its path, is $\sim$~\SI{100}{\femto\second}, where the electrons continue to be excited, until the ground state is depleted, leading to transient bleaching with saturated absorption $\Delta A < 0$.
The bleaching occurs around \SI{1}{\pico\second} (\SI{1}{\tera\hertz}) and recovers within \SI{2}{\pico\second}, which can be used as an all-optical switch \cite{Pianelli2024} and saturable-absorption-enabled hybrid mode-locking in a laser with an intracavity ENZ component \cite{Wu2022lpr, Wu2025}.
After this process, at $t \approx 3$ \si{\pico\second} the thermo-optic effects becomes prominent as electron-phonon (e-ph) interaction and phonon scattering heat up the lattice, peaking at $\sim$~\SI{8}{\pico\second} (Fig.~\ref{f3}(c), dark orange line).
This near-THz thermo-optic response is on par with the predicted values for nanoparticles \cite{Khurgin2015} and much faster than the $\sim$~\SI{10}{\mega\hertz}-grade response in gallium phosphide integrated photonic chip \cite{Wilson2020}, silicon nanocavities \cite{Clementi2021}, and silicon nitride microresonators \cite{Liu2021} (see Supplementary Table 3).
The thermo-optic nonlinearity response increases with pump power and saturates at $\sim$~\SI{160}{\micro\watt}.
The measured $1/e~\Delta A$ intensity fall time is around \SI{100}{\pico\second}, which is on par with the rise time of many conventional homogeneous materials \cite{Boyd2020}, while limiting the overall switching and modulation speed.

To perform a qualitative analysis, we analyze the following two-temperature model (TTM) \cite{Sun1993,Alam2016, Xie2022} by treating the effective medium as a whole:

\begin{equation}
    \begin{array}{l}
C_{\mathrm{e}} \, \dfrac{\partial T_{\mathrm{e}}}{\partial t}
= -G_{\mathrm{e\text{-}ph}} (T_{\mathrm{e}} - T_{\mathrm{lattice}}) + \alpha I(t), \\
C_{\mathrm{lattice}} \, \dfrac{\partial T_{\mathrm{lattice}}}{\partial t}
= G_{\mathrm{e\text{-}ph}} (T_{\mathrm{e}} - T_{\mathrm{lattice}}) -\kappa_{\mathrm{eff}}\nabla^2 T_{\mathrm{lattice}},
\end{array}
    \label{e5}
\end{equation}

\begin{figure*}[!h]
	\centering
	\includegraphics[width=1\linewidth]{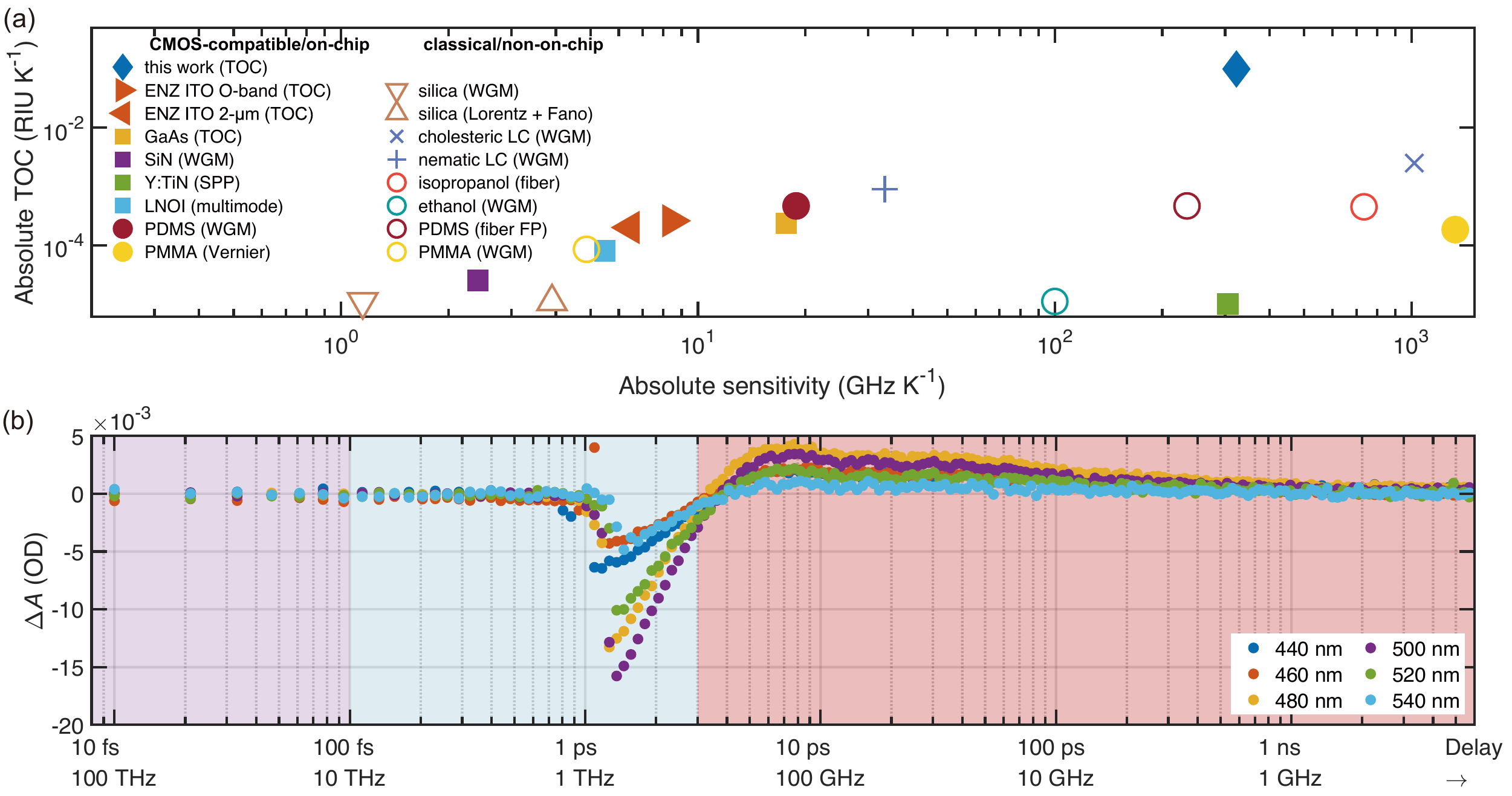} 
	\caption{{. Epsilon-near-zero metamaterial for ultrasensitive and transient temperature detection.} 
    (a) Performance comparison with state-of-the-art: the absolute values of TOC and sensitivity.
    In subfigure (b), the solid markers denote materials with CMOS-compatible or on-chip implementations, while the hollow markers denote those without.
    In the left column of the legend, the data sources are: ENZ ITO \cite{Wu2024}, GaAs \cite{chenOnchipHighsensitivityPhotonic2020}, SiN \cite{zhangPhotonicThermometerSilicon2022}, Y:TiN \cite{rahmanHighlySensitiveTemperature2026}, LNOI \cite{wangEnhancedTemperatureSensing2024}, PDMS \cite{liChipHighsensitivityThermal2010}, PMMA \cite{tianTemperatureSensorHighsensitivity2020}.
    In the right column of the legend, the data sources are: silica (WGM) \cite{gaoUltrahighresolutionHighorderWGM2025}, silica (Lorentz + Fano resonances) \cite{yangTemperatureSensingBased2025}, cholesteric LC \cite{zhaoWhisperingGalleryMode2017}, nemanic LC \cite{kavungalThermoopticTuningPackaged2018}, isopropanol \cite{zhaoHighlySensitiveTemperature2018}, ethanol \cite{wardHighlySensitiveTemperature2013}, PDMS \cite{dongFabricationHighQPolydimethylsiloxane2009}, PMMA \cite{heTemperatureSensorBased2018}.
    The values of the parameters are available in Supplementary Table 2.
    (b) Ultrafast dynamics of the whole processes in the pump-probe light-matter interactions at selected wavelengths near $\lambda_{\rm ENZ}$
    }
	\label{f5}
\end{figure*}

\noindent where $C_{\rm e}$ and $C_{\rm lattice}$ are the heat capacities of the electron and lattice, $T_{\rm e}$ and $T_{\rm lattice}$ are their respective effective temperatures. 
$G_{\rm{e\text{-}ph}}$ is the electron-phonon coupling coefficient.
Under the ENZ conditions, the electric field is enhanced by the boundary continuity \cite{Campione2013} of the electric field.
The absorption $\alpha$ is also enhanced by ENZ, as illustrated in Fig~\ref{f2}(b) in red. 
Therefore, the source term $\alpha I$ of the first equation in Eq.~\eqref{e5} becomes larger.
Upon ultrafast laser irradiation, the $\gamma$-e and e-e interactions are ultrafast (atto- to femtosecond), $T_{\rm e}$ is expected to heat up dramatically.
Since the rising of $T_{\rm lattice}$ relies on the much slower (at least larger than picosecond-scale) e-ph coupling, at this stage, $T_{\rm lattice}$ remains almost unchanged.
Although no ENZ-induced acceleration occurs in the first stage of interaction, a temperature gradient appears for the e-ph interactions. 
Considering the second equation of Eq.~\eqref{e5}, this results in an increase of $\partial T_{\rm lattice}/\partial t$.
Hence, the ENZ condition dramatically enhances the magnitude of the initial lattice heating rate, leading to a much stronger and faster-rising thermo-optic response.
The dissipation term $\kappa_{\rm eff}\nabla^2 T_{\rm lattice}$ is negligible compared to the accelerated thermo-optic response, when the lattice is still ``cold''.
This term, however, governs the relaxation time of the thermo-optic effects at a much slower rate $\kappa_{\rm eff}/C_{\rm lattice}$.
During the whole process, the effective ENZ medium is being dynamically reconfigured in time by the thermo-optic responses $\varepsilon[T(t)]$ and thermal expansion responses $d[T(t)]$ of the constituent materials, while maintaining the ENZ condition that supports the behavior near the original $\nu_{\rm ENZ}$.

Notably, the observed enhancement and acceleration of thermo-optic nonlinearity can be attributed to a dynamic time-varying thermo-optic reconfiguration of the effective ENZ medium.
A quantitative analysis using TTM simulation is performed using Eq.~\eqref{e5}, and the results are shown in Fig.~\ref{f4}.

In Fig.~\ref{f4}(a), $T_{\rm e}$ peaks almost instantaneously after ultrafast pulse stimulation.
It drops to half of its maximum intensity at \SI{3.2}{\pico\second} where the $T_{\rm lattice}$ begins to dominate $\Delta A$ variation and thermo-optic nonlinearity becomes clearly visible in Figs.~\ref{f3}(c) and (d).
$T_{\rm lattice}$ reaches its maximum at \SI{8.1}{\pico\second} which agrees well with experiments.
The plasma frequency $\omega_{\rm p}$ of the silver nanolayers \cite{Ferrera2019}, characterized by Drude model \cite{Drude1900}, also shows a saturation at \SI{8.1}{\pico\second}.
Similar to the static reconfiguration of the effective medium in Fig.~\ref{f1}(c), the temporal evolution of refractive indices of the effective medium components, silver (Fig.~\ref{f4}(c)) and silica (Fig.~\ref{f4}(d)) dynamically reconfigures the effective ENZ medium in time, resulting in the thermo-optic response profile shown in Fig.~\ref{f3}.
A more refined simulation approach is possible \cite{Sarkar2023} but goes beyond the scope of the work.

One might argue that the reported behavior is dominated only by the silver nanolayer (Figs.~\ref{f4}(b) and (c)), and the almost-unchanged silica seems to have no effect (Fig.~\ref{f4}(d)).
However, as can be seen from Figs.~\ref{f3}(b), (c), and~\ref{f5}(b), the thermo-optic nonlinearity is enhanced near $\lambda_{\rm ENZ}$ (the \num{480} and \SI{500}{\nano\metre} curves) which is not found in a standalone silver nanolayer.

\section{Discussion}

\subsection{Novel and ultra-sensitive temperature detection mechanism}

Modern consumer electronics and commercial computational end-products operate under the maximum junction temperature to avoid functional failure \cite{Chen2008}.
For emerging applications and industrialization of PICs, the thermal constraints are similar, especially when they are working closely and collaboratively with electronic integrated chips (EICs).
Some PICs, due to strict resonance and coupling conditions, require even more precise and faster-response temperature monitoring and controls \cite{Wilson2020,Wu2023}.
An ideal temperature sensor in this scenario should be robust, structurally simple, highly sensitive, and ready for deployment in large volume on multiple hotspots inside a LSI chip with low cost and relatively small surface area.

State-of-the-art photonic temperature sensors exploit the thermo-optical properties of organic materials such as polydimethylsiloxane (PDMS) \cite{yaoSingleOpticalMicrofiber2022, zhangAllfiberTemperatureHumidity2023, dongFabricationHighQPolydimethylsiloxane2009, liChipHighsensitivityThermal2010, liangImprovementTemperatureSensitivity2024, wangTemperatureSensorSinglemodenocoresinglemode2022}, poly(methyl methacrylate) (PMMA) \cite{tianTemperatureSensorHighsensitivity2020, heTemperatureSensorBased2018}, ethanol \cite{guHighlySensitiveTemperature2019, wardHighlySensitiveTemperature2013}, isopropanol \cite{zhaoHighlySensitiveTemperature2018}, and aceton \cite{liuDualSensitizationEnhancement2024}, liquid crystals (LCs) \cite{zhaoWhisperingGalleryMode2017, kavungalThermoopticTuningPackaged2018}, as well as inorganic materials like gallium arsenide (GaAs) \cite{chenOnchipHighsensitivityPhotonic2020}, yttrium-doped titanium nitride (Y:TiN) \cite{rahmanHighlySensitiveTemperature2026}, silicon nitride (SiN)\cite{zhangPhotonicThermometerSilicon2022}, silica (SiO$_2$) \cite{yangTemperatureSensingBased2025, gaoUltrahighresolutionHighorderWGM2025}, lithium niobate on insulator (LNOI) \cite{wangEnhancedTemperatureSensing2024}, molybdenum disulfide (MoS$_2$) \cite{mohanrajAllFiberopticUltrasensitive2018}, \textit{etc.}
They take full advantages of various effects to translate temperature change into a frequency shift, including whispering gallery mode (WGM) resonators \cite{dongFabricationHighQPolydimethylsiloxane2009, liChipHighsensitivityThermal2010}, thermo-optic coefficient (TOC) \cite{chenOnchipHighsensitivityPhotonic2020}, Fabry-Pérot (FP) cavity \cite{zhangAllfiberTemperatureHumidity2023}, surface plasmon ploaritons (SPP) \cite{rahmanHighlySensitiveTemperature2026}, Lorentz and Fano resonance \cite{yangTemperatureSensingBased2025}, Vernier effect \cite{tianTemperatureSensorHighsensitivity2020}, \textit{etc.}
Organic materials like the cholesteric LC in particular, can reach a high sensitivity of over \SI{1000}{\giga\hertz\per\kelvin} with a large TOC at the level of \SI{1e-3}{\riu\per\kelvin} \cite{zhaoWhisperingGalleryMode2017}.
Although many of the implementations are on-chip or CMOS-compatible \cite{liChipHighsensitivityThermal2010, chenOnchipHighsensitivityPhotonic2020, wangEnhancedTemperatureSensing2024,yangTemperatureSensingBased2025,rahmanHighlySensitiveTemperature2026}, they face challenges of compositional robustness (PDMS/PMMA/LCs), low TOC (GaAs, SiN, LNOI, SiO$_2$), relying on waveguide coupling conditions, and low overall sensitivity (prominent in inorganic materials even with WGM).

Here, without any resonant structure and occupying a relatively small unpatterned die area as compared to ring resonators, the proposed metamaterial in Fig.~\ref{f1}(a) can serve as a standalone unit for ultrasensitive temperature detection.
By only relying on the raw data of reflected light (Eq.~\ref{e1}), $\nu_{\rm ENZ}$ can be extracted and used for ultrasensitive temperature sensing, which is an unreported novel approach.
It is worth emphasizing that the metamaterial itself is already at the deployable device level when it is used as an optical temperature sensor, and the thermal-spectral modulation rate shown in Fig.~\ref{f1}(b) can directly translate into temperature sensitivity (\SI{322.7}{\giga\hertz\per\kelvin}).

In the comparison presented in Fig.~\ref{f5}(a), the sensitivity of the proposed effective ENZ metamaterial exceeds all known inorganic CMOS-compatible materials with or without resonance cavities, and is on par with the organic on-chip PMMA sensor based on the Vernier effect, potentially enabling more performant nanosensors.
This result is verified by performing multiple repeated measurements at different $\Delta T$ (see Supplementary Notes 5 and 7).
Such unique feature of the metamaterial can be used as a competitive alternative to organic sensing media.

Resembling the conventional on-chip temperature sensing device like the WGM PIC, its temperature readouts and analyses rely on external equipments.
For practical implementation, any compact or integrated optical measurement scheme \cite{yangSinglenanowireSpectrometers2019} capable of probing the ENZ spectral position, such as polarization-resolved reflection analyses or transmission/reflection spectroscopy, could be employed to realize the proposed ultrasensitive thermo-optic sensing.
Detailed comparison tables can be found in Supplementary Note 8 \cite{Wu2024, chenOnchipHighsensitivityPhotonic2020, zhangPhotonicThermometerSilicon2022, rahmanHighlySensitiveTemperature2026, wangEnhancedTemperatureSensing2024, liChipHighsensitivityThermal2010, tianTemperatureSensorHighsensitivity2020, gaoUltrahighresolutionHighorderWGM2025, yangTemperatureSensingBased2025, zhaoWhisperingGalleryMode2017, kavungalThermoopticTuningPackaged2018, zhaoHighlySensitiveTemperature2018, wardHighlySensitiveTemperature2013, dongFabricationHighQPolydimethylsiloxane2009, heTemperatureSensorBased2018, Liu2021, hbeebThermoopticCoefficientLithium2024, dongFabricationHighQPolydimethylsiloxane2009, gaoInvestigationThermoOpticCoefficient2018a, Suresh2021, Alam2016, Huang2023, Clementi2021, Duh2020, Yan2017, Leuthold2010a, Liu2021, Wilson2020}. 

\subsection{Transient detection of a femtosecond-pulse-induced heat source}

Owing to the enhanced and accelerated thermo-optic response, the effective ENZ metamaterial holds promise for interesting and novel applications, such as the detection of a transient heat induced by a femtosecond laser pulse.
This feature is especially useful in a packaged photonic system-on-chip to capture the breaking of photonic circuits and the leakage of laser pulses as part of risk control and management.

The pump-probe experiment in Fig.~\ref{f3}, in the context, emulates the leakage of a sheer amount of ultrafast heat in a broken LSI PIC product.
In each test cycle, only one 100-fs laser pulse is launched, carrying an energy of \SI{100}{\nano\joule} per pulse under \SI{100}{\micro\watt} average power.
This amount of energy can cause a transient temperature rise of \SI{0.23}{\kelvin} (see Supplementary Note 4 for estimation), which is hard to detect by conventional photonic temperature sensors.

Figure~\ref{f5}(b) illustrates the sensitivities of transient heat detection of different wavelengths near $\lambda_{\rm ENZ}$.
An ENZ-induced enhancement can be found by observing the \num{480} and \SI{500}{\nano\metre} curves.
The overall sensitivity is \num{3.1e-3}~OD~K$^{-1}$ or \SI{-3e-2}{\decibel\per\kelvin} (see Supplementary Note 4).
Any pulse energy larger than that is still detectable but will not be captured in the form of a higher $\Delta T$ or $\Delta A$ (OD).

In practical deployment, a weak probe light can be shone on the nanosensor's surface and monitor the change in absorption.
Note that the second nonlinear stage, bleaching, is less suitable for ultrafast pulse detection since it requires sufficiently high energy to trigger ground-state depletion; while the thermo-optic effects exist as long as the driving force, $\Delta T$ exists.
The broadband nature of the ENZ-enhanced thermo-optic nonlinearity response allows the detection of a wider range of pulse wavelengths without the need to design/deploy sources for a specific wavelength.

It is worth noting that while bare silver is generally restricted in standard front-end-of-line (FEOL) CMOS manufacturing, our stratified architecture inherently encapsulates the Ag films within alternating SiO$_2$ dielectric layers, readily lending itself to back-end-of-line (BEOL) integration schemes. 
Furthermore, the underlying ENZ-enhanced thermo-optic mechanism is fundamentally material-agnostic. 
For strict fabrication compatibility in future on-chip applications, the constituent Ag layers could be substituted with fully CMOS-compatible metals, such as aluminum (Al) or copper (Cu), tailored for specific operating wavelength regimes. 
This material interchangeability substantially broadens the general applicability of our proposed thermo-optic framework.

\section{Conclusions}

In this work, we establish thermo-optic reconfiguration of effective media as a unifying physical framework for understanding both static and dynamic thermo-optic phenomena in epsilon-near-zero photonic systems.
Using a metal-dielectric effective medium, we experimentally demonstrate that temperature variation, whether under thermal equilibrium or transient excitation, reconfigures the constitutive parameters defining the ENZ condition, giving rise to pronounced linear and nonlinear optical responses.
This reconfiguration manifests as a large $\nu_{\rm ENZ}$ shift, and as a picosecond-scale thermo-optic nonlinear response in time-dependent ENZ interfaces.

Beyond the specific material platform and experimental configuration studied here, the concept of thermo-optic reconfiguration highlights a general mechanism by which thermal, structural, and optical degrees of freedom are intrinsically coupled in effective photonic media.
By bridging effective-medium theory, thermo-optic physics, and time-varying photonics, this work also offer further insight into ENZ nanodevices, position ENZ nanostructures as universal thermo-optical mediators, open routes for real-time heat-flow mapping, ultrafast thermal logic, and neuromorphic photonics, and establish ENZ materials as a viable platform for exploiting diverse nonlinear optical phenomena.

\section*{Disclosures}

{\small The authors declare no competing interests.}

\section*{Data availability}

All data needed to evaluate the conclusions in the paper are present in the paper and the Supplementary Information.

\section*{\textit{Acknowledgements}}

{\small 
Z.-K.Z. acknowledges support by the National Key Research and Development Program of China (2021YFA1400804) and the National Natural Science Foundation of China (12222415).
J.W. acknowledges support by the City University of Hong Kong (Dongguan) Startup Fund.
M.C. acknowledges support by the Italian Ministry of Education (MUR) PNRR project PE0000023-NQSTI.
}

\begingroup
\scriptsize	
\bibliography{References}
\endgroup

\end{document}